\documentclass[%
  preprint,
  superscriptaddress,
  amsmath,amssymb,
  aps,
  pre
]{revtex4}

\usepackage{mathtools}
\usepackage{color}
\usepackage{amsmath}
\usepackage{graphicx}
\usepackage{amssymb}
\usepackage{amsthm}
\usepackage{upgreek}

\begin{document}

\title{Optimal subgrid scheme for  shell models of turbulence}

\author{Luca Biferale}
\affiliation{Dept. Physics and INFN,
  University of Rome ``Tor Vergata", Via della Ricerca Scientifica 1,
  I-00133 Roma, Italy.}
  
\author{Alexei A. Mailybaev}
\affiliation{Instituto Nacional de Matem\'atica Pura e Aplicada -- IMPA, 
Estrada Dona Castorina 110, 22460--320 Rio de 
Janeiro, Brazil. }

\author{Giorgio Parisi}
\affiliation{Dept. Physics and INFN,  University of Rome 'Sapienza', Piazzale A. Moro 5, I-00185 Rome,  Italy. }

\begin{abstract}
We discuss a theoretical framework to define an optimal sub-grid closure for shell models of turbulence. The closure is based on the ansatz that consecutive  shell {\it multipliers} are short-range correlated, following the third hypothesis of Kolmogorov formulated for similar quantities for the original  three-dimensional Navier--Stokes  turbulence. We also propose a series of systematic approximations to the optimal model by assuming different degrees of correlations across scales among amplitudes and phases of consecutive multipliers. We show numerically that such low-order closures work well, reproducing all known properties of the large-scale dynamics including anomalous scaling. We found small but systematic discrepancies only for a range of scales close to the sub-grid threshold, which do not tend to disappear by increasing the order of the approximation. We speculate that the lack of convergence might be due to a structural instability, at least for the evolution of very fast degrees of freedom at small scales. Connections with similar problems for Large Eddy Simulations of the three-dimensional Navier--Stokes equations are also discussed. \\ {\bf Postprint version of the article published on Phys. Rev. E 95, 043108 (2017) DOI: 10.1103/PhysRevE.00.003100}
\end{abstract}

\maketitle

\section{Introduction}
Three-dimensional turbulence is a multiscale phenomenon triggered  when 
the nonlinear transport terms in the Navier--Stokes (NS)  equations 
 are much more intense than the viscous linear damping \cite{Fri97}. 
The control parameter is given by the Reynolds number, $Re= u_0l_0/\nu$, made out of  the 
typical root mean square velocity, $u_0$, the typical length scale, $l_0$ and the 
kinematic viscosity, $\nu$. It is an empirical fact that in the turbulent regime the flow 
develops a  {\it dissipative anomaly}: a $Re$-independent energy transfer, from the scale where the external forcing  is acting till the viscous range.
The energy transfer mechanism is characterized by anomalous scaling laws and by 
a highly non-Gaussian and intermittent statistics \cite{Fri97}. It is fair to say that we
 do not yet possess neither the analytical nor the numerical tools 
to fully  quantify  turbulence  for  three-dimensional flows. 
 
Shell models provide a natural playground for fundamental studies of developed turbulence \cite{Fri97,bohr,biferale2003shell,ditlevsen2010turbulence}. These models allow accurate numerical simulations and possess nontrivial properties of the Kolmogorov--Obukhov theory for turbulence at high Reynolds numbers: a forward energy transfer, a dissipative anomaly and intermittency with  
anomalous scaling similar to what observed for the original three-dimensional NS equations.  The idea is to build simple models  
 sharing the key statistical properties of the turbulent energy cascade. In this paper, we focus on the Sabra shell model~\cite{l1998improved} (a modified version of the Gledzer--Ohkitani--Yamada model~\cite{gledzer1973system,ohkitani1989temporal,bohr}), which is obtained by reducing dynamics to a discrete sequence of shells $|\mathbf{k}| = k_n$ in the Fourier space for the geometric 
progression of wavenumbers $k_n = k_0\lambda^n$, $n = 1,2,3,\ldots$ (we use $k_0 = 1$ and $\lambda = 2$).  The turbulent ``flow'' is described by complex velocity variables $u_n(t)$, which mimic the velocity increments at the corresponding shells, $u_n \sim \delta_\ell v = v(\ell)-v(0)$. Thus, the shell variable $u_n$ characterizes the velocity fluctuation at  scale $\ell \sim 1/k_n$. 

One of the main theoretical and applied challenges in the theory of turbulence consists in {\it 
closing} the  equations of motion on a coarser grid, i.e. to derive a model for the small-scale degrees of freedom to be used to evolve the variables at large scales. 
The problem is key for Large Eddy Simulations (LES),  a set of  applied 
numerical tools meant to reduce the computational costs to simulate high Reynolds number turbulence \cite{meneveau2000scale, pope2001turbulent, sagaut2006large, les87}. The problem is also key from a theoretical point of view, because, if successful, would imply a complete control on 
the energy-transfer mechanism at all scales.
The main difficulties to accomplish the goal  for the three-dimensional  NS case are connected to 
the extremely complicated functional and statistical dependency of the unresolved sub-grid variables from the resolved ones, the legacy of the strong non-linear character of the dynamical evolution together with the strong non-local coupling  in both  real and  Fourier space of the original equations. In fact, despite many advancements, the problem of finding an optimal sub-grid model to be applied in LES is considered still open. 

Our aim here is to show that this task can be accomplished for the Sabra shell model in a way that accurately describes the statistics of subgrid scales. The good news  is that the simplified structure
 of the non-linear terms  allows for a precise theoretical and numerical analysis of the statistical coupling among resolved and unresolved shells. As a result, it is possible to define what would be the {\it optimal} closure, in theory.   The bad news is that the problem is not of easy implementation  even in this case and that it is difficult to figure out a systematic protocol of more and more complex sub-grid models which converge toward the ``optimal'' 
one.  The main idea is to close the sub-grid dynamics in terms of
 multi-scale correlations among {\it multipliers}, i.e. ratios among consecutive  shell variables \cite{benzi1993intermittency,eyink2003gibbsian,benzi2004,friedrich2016generalized,renner2001experimental,ragwitz2001indispensable,schmitt2003causal}. 
The approach goes back to the third hypothesis of Kolmogorov \cite{Fri97}, 
made to disentangle universal 
small-scale  fluctuations from non-universal coupling with the large-scale motion. 
Differently from  the original case of NS equations, multipliers in shell models follow a simple
 non-linear dynamical  evolution. It is therefore possible to manipulate them and to make predictions \cite{benzi1993intermittency,eyink2003gibbsian}. It turns out that it is crucial to distinguish the correlations among their amplitude and their
 phases. In this paper we first  show how to define a  formal optimal sub-grid model. The model is still too complicated to be implemented in practice, being defined in terms of the conditional probability of a few 
 sub-grid variables with {\it all} resolved degrees of freedom, a task out of 
reach even for simplified dynamics as for the case of shell models. Then, we show how to 
 develop a series of simple approximations  for the sub-grid closure 
that work well, i.e. they are able to quantitatively 
 reproduce the large-scale dynamics except for a short range of shells close to the
 cutoff. We also show that the observed deviations are Reynolds independent, i.e. the discrepancies remain 
localized to a limited number of scales close to the cut-off independently of the intensity  of turbulence.
Unfortunately, numerics demonstrates that the proposed systematic protocol of more and more refined closures denies a controllable convergence to the optimal model at small scales. We speculate that this might be due 
to non-trivial strong sensitivity of the structure of the attractor on the small-scale closure, a sort of breaking of ergodicity at fast small-scale degrees-of-freedom. A comment 
on the  potential connections with the equivalent  problem to find an optimal sub-grid closure for LES of turbulence is also proposed. 

The paper is organized as follows. In Sec.~\ref{sec3} we discuss the  set-up on how to define the optimal 
sub-grid model for a general shell model. In Sec.~\ref{sec4} we show how to implement the third hypothesis of Kolmogorov to define a systematic universal closure for the sub-grid model. In Sec.~\ref{sec4b} 
 we show how this procedure works in simple shell models  where the dynamical evolution is not intermittent. In Sec.~\ref{sec6} we formulate it for the case of the Sabra model, one of the most popular and studied shell models for turbulence. 
In the same section, we propose and apply a set of approximations to the optimal closure  for the Sabra model and  discuss
their pluses and minuses. Conclusions follow in Sec.~\ref{sec:conc}.

\section{Reduced system for a probability density}
\label{sec3}
Shell models are  dynamical systems 
 which mimic the fluid dynamics by considering 
a geometric progression of wavenumbers, $k_n = k_0\lambda^n$, for some fixed $\lambda > 1$ and $n = 1,2,\ldots,N$. Each wavenumber defines a shell $|k| = k_n$ in Fourier space represented by one or several shell variables, which describe intensity of the flow at a corresponding scale. Characteristic scale in physical space can be defined as $\ell \sim 1/k_n$. Thus, $n \sim 1$ corresponds to large scales $\ell \sim 1/k_0$, while $n \sim N$ yields the smallest scales of the system.

For simplicity, we start by assuming real shell variables $u_n$ and considering a model with only the nearest-shell interaction. These assumptions are made in order to present the derivations in a simple and clear form, and then we 
extend the results to general shell models in Sec. 5. Equations of our simple shell model read
	\begin{equation}
	\dot u_n = k_nQ_n-\nu k_n^2 u_n, \quad n = 1,\ldots,N,
	\label{eqE0}
	\end{equation}
with  the quadratic nonlinear term coupling only the nearest neighbors: 
	\begin{equation}
	Q_n = Q(u_{n-1},u_n,u_{n+1}) = \sum_{i,j \in \{-1,0,1\}} a_{ij}u_{n+i}u_{n+j}.
	\label{eqE0b}
	\end{equation}
A boundary condition must be supplied for the initial shell
	\begin{equation}
	u_0 = u_0(t).
	\label{eqE0c}
	\end{equation}
The total number of shells $N$ is assumed to be large enough leading to the strong decay due to viscosity at small scales, i.e., $u_N \approx 0$.
Note that we use no explicit forcing term in Eq.~(\ref{eqE0}), with the excitation performed by the boundary condition (\ref{eqE0c}) as it is typical for realistic flows. The nonlinear term in (\ref{eqE0b}) must be chosen such that the system possesses an inviscid invariant $E = \frac{1}{2}\sum u_n^2$ called the energy.
	
The number of shells involved in the dynamics depends on viscosity $\nu$. Considering the integral scales of the system $L \sim 1/k_0 \sim 1$ and $T \sim 1$, the Reynolds number is defined simply as $\mathrm{Re} = 1/\nu$. In statistically stationary regime with large Reynolds numbers, one can
 distinguish three ranges of scales with qualitatively different behavior~\cite{Fri97}. The range of large scales, $n \sim 1$, is called the forcing range, as it is influenced by the boundary conditions producing the energy input into the system. The energy dissipates at small scales $n \gtrsim n_K$ of the viscous range. The estimate 
	\begin{equation}
	n_K \approx -\frac{3}{4}\log_\lambda \nu
	\label{eqK1}
	\end{equation}
can be obtained by comparing $\ell \sim 1/k_n$ with the Kolmogorov scale $\eta = (\nu^3/\varepsilon)^{1/4}$, where $\varepsilon \sim 1$ is the rate of energy dissipation~\cite{Fri97}. For large Reynolds numbers (small viscosity) the forcing range, where energy is injected, is separated from the viscous range, where it dissipates. The intermediate range with $L \gg 1/k_n \gg \eta$ is called the inertial interval. In the inertial interval, both forcing and viscosity can be neglected leading to a positive mean energy flux $\varepsilon$ from larger to smaller scales, called the energy cascade. 

We will consider the evolution of a statistical ensemble, corresponding to some probability distribution as initial condition.	 We denote by $P(u_1,\ldots,u_N;t)$ a probability density of the shell variables at time $t$. Time dependence of this distribution is governed by the continuity equation
	\begin{equation}
	\frac{\partial P}{\partial t} 
	+ \sum_{n = 1}^N 
	\frac{\partial }{\partial u_n} \left(\dot u_n P\right) = 0.
	\label{eqE1}
	\end{equation}	
Our goal is to derive a reduced model for a given sequence of  shells variables, $u_1,\ldots,u_s$, where $s$ is any shell number from the inertial interval. The latter means that the viscous term in Eq.~(\ref{eqE0}) can be neglected for the corresponding shells with
	\begin{equation}
	\dot u_n = k_nQ_n,\quad n = 1,\ldots,s.
	\label{eqE2}
	\end{equation}
The  reduced probability distribution is defined as the result of integration over all shells with $n>s$:
	\begin{equation}
	P_s(u_1,\ldots,u_s;t) = \int P(u_1,\ldots,u_N;t) \prod_{m = s+1}^N du_m.
	\label{eqE3}
	\end{equation}
Similar integration applied to Eq.~(\ref{eqE1}) yields
	\begin{equation}
	\frac{\partial P_s}{\partial t} 
	+ \sum_{n = 1}^N 
	\int \frac{\partial }{\partial u_n} \left(\dot u_n P\right)\prod_{m = s+1}^N du_m = 0.
	\label{eqE4}
	\end{equation}
The terms with the derivatives $\partial/\partial u_n$ for $n = s+1,\ldots,N$ vanish after the integration with respect to $u_n$. For the other terms, we write $P = (P/P_s)P_s$ and substitute in (\ref{eqE2}). The resulting equation becomes
	\begin{equation}
	\frac{\partial P_s}{\partial t} 
	+ \sum_{n = 1}^s
	\frac{\partial }{\partial u_n} \left(k_nR_n P_s\right) = 0,
	\label{eqE5}
	\end{equation}
where
	\begin{equation}
	R_n(u_1,\ldots,u_s;t) = 
	\int Q_n \frac{P}{P_s} \prod_{m = s+1}^N du_m.
	\label{eqE6}
	\end{equation}
Here we specified that the functions $R_n$ may depend on all shell variables $u_1,\ldots,u_s$ and time, due to the corresponding dependence of $P$ and $P_s$. 
The key point is to realize that equation (\ref{eqE5}) describes the evolution of the probability density for a reduced  dynamical system:
	\begin{equation}
	\dot u_n = k_nR_n,\quad n = 1,\ldots,s.
	\label{eqE7}
	\end{equation}
Eq.~(\ref{eqE7}) will be our coarse grained  system, when we obtain closed expressions for the right-hand sides in (\ref{eqE6}) as functions of the variables $u_1,\ldots,u_s$.

The very same approach  can be followed for the full Navier--Stokes equations, see \cite[Chap.~13.5.6]{pope2001turbulent}. 
The main advantage given by  shell models is that they have only 
local or quasi-local interactions among consecutive shells. Indeed,  the factor $Q_n = Q(u_{n-1},u_n,u_{n+1})$ does not depend on the integration variables in (\ref{eqE6}) for those shells with  $n < s$, while $Q_s = Q(u_{s-1},u_s,u_{s+1})$ depends on $u_{s+1}$. Hence, the integration in (\ref{eqE6}) can be carried out using (\ref{eqE3}) and leading to the explicit  expressions
	\begin{equation}
	R_n = \left\{\begin{array}{ll}
	Q_n, & n = 1,\ldots,s-1;\\[7pt]
	\displaystyle
	\int Q_s \frac{P_{s+1}}{P_s} \,du_{s+1},& n = s.
	\end{array}\right.
	\label{eqE8}
	\end{equation}
Here $P_{s+1}(u_1,\ldots,u_{s+1};t)$ is defined by the expression analogous to (\ref{eqE3}). We see that the original system (\ref{eqE2}) and the reduced system (\ref{eqE7}), (\ref{eqE8}) differ only by the last equation. This is natural because the nonlinear term $Q_s$ is the only one that depends on the unknown shell variable $u_{s+1}$. Thus, the only missing component of the reduced system is the unknown integral expression in Eq.~(\ref{eqE8}). In the jargon of LES the sub-grid model is influencing the explicit dynamical evolution of only one resolved variable (but still depends on the correlations with all of them).  

In general, one needs to know the whole distribution $P_{s+1}(u_1,\ldots,u_{s+1},t)$ to compute $R_s$ in (\ref{eqE8}). The main idea of this paper is that the form of the function $R_s$ is in fact universal in the developed turbulent dynamics, as suggested by numerical simulations and some theoretical considerations described below. This observation is central for our work and provides the subgrid model (\ref{eqE8}) in closed form.

\section{Kolmogorov's third hypothesis and universality of the reduced equations}
\label{sec4}
In 1962, Kolmogorov~\cite{kolmogorov1962refinement} conjectured that the statistics of velocity increment 
 ratios (multipliers) $\delta_\ell v / \delta_{\ell'} v$ is universal and depends only on the
 scale ratio $\ell/\ell'$ in the inertial interval of homogeneous isotropic hydrodynamic turbulence. This conjecture, called the third Kolmogorov hypothesis, 
was confirmed both numerically and 
experimentally~\cite{Fri97,chen2003kolmogorov,lvo96a,ben98,chhabra1992scale,pedrizzetti1996self,jouault1999multiplier,nelkin1996limitations}.
 For shell models, this conjecture implies that the probability distribution 
of multipliers $z_{n} = u_n/u_{n-1}$
is universal and does not depend on $n$ in the inertial interval, which agrees very well with numerical simulations for the Sabra shell model~\cite{eyink2003gibbsian}. Furthermore, Kolmogorov assumed that the multipliers for widely separated shells are statistically independent. Indeed, the distribution of multipliers appears to be short-range, i.e., correlations between $z_n$ and $z_{n+j}$ decay rapidly with increasing $j$. 

The factor in the integral expression (\ref{eqE8}),
	\begin{equation}
	\frac{P_{s+1}}{P_s} \,du_{s+1} = P_{\textrm{cond}}(u_{s+1}|u_s,\ldots,u_1;t) du_{s+1},
	\label{eqE9}
	\end{equation}
is by definition the conditional probability of $u_{s+1}$ for given $u_s,\ldots,u_1$ at time $t$. Note that there is a one-to-one correspondence between the shell variables $u_1,\ldots,u_s$ (with given boundary condition for $u_0$) and the multipliers $z_1,\ldots,z_s$ in the case when all of them are nonzero. The singular subset, when one of the variables vanishes, has zero measure and it is not important for our probabilistic analysis. 
Similarly, there is one-to-one correspondence between the shell variable $u_{s+1}$ and the multiplier $z_{s+1}$ for given $u_1,\ldots,u_n$ (or $z_1,\ldots,z_n$). Hence,
the change of variables from $u_n$ to $z_n = u_n/u_{n-1}$ yields the conditional probability for shell variables in terms of the conditional probability for multipliers as 
	\begin{equation}
	P_{\textrm{cond}}(u_{s+1}|u_s,\ldots,u_1;t) du_{s+1}
	= \widetilde{P}_{\textrm{cond}}(z_{s+1}|z_s,\ldots,z_1;t) dz_{s+1}.
	\label{eqE10}
	\end{equation}
The third Kolmogorov hypothesis for the developed turbulent regime implies that the function
	\begin{equation}
	\widetilde{P}_{\textrm{cond}}(z_{s+1}|z_s,\ldots,z_1;t)
	=  \widetilde{P}_{\textrm{uni}}(z_{s+1}|z_s,z_{s-1},\ldots)
	\label{eqE10b}
	\end{equation}
is universal and time-independent, such that it is uniquely determined for a given shell model. Also, $\widetilde{P}_{\textrm{uni}}$ must have short-range dependence on its arguments, i.e., it depends essentially only on a few neighboring shells $z_s,z_{s-1},\ldots$ with very weak dependence on $z_n$ for smaller $n$. From now on, we will use the arguments written as $z_s,z_{s-1},\ldots$ to indicate such a short-range dependence.

Note that Eqs.~(\ref{eqE10}) and (\ref{eqE10b}) do not necessarily imply the universality of $P_{\textrm{cond}}(u_{s+1}|u_s,\ldots,u_1;t)$. This is because the correlations between the shell variables $u_n$ extend to the whole range of scales and, hence, $P_{\textrm{cond}}$ may depend on the first (large scale) shells and on the boundary conditions.

Using Eqs.~(\ref{eqE9})--(\ref{eqE10b}) and (\ref{eqE0b}) in (\ref{eqE8}), yields
	\begin{equation}
	R_s(u_s,u_{s-1},\ldots) 
	= \int Q(u_{s-1},u_s,u_{s+1}) \widetilde{P}_{\textrm{uni}}(z_{s+1}|z_s,z_{s-1},\ldots) dz_{s+1}.
	\label{eqE12a}
	\end{equation}
Since $Q$ in (\ref{eqE0b}) is a quadratic function of its arguments, we can write
	\begin{equation}
	Q(u_{s-1},u_s,u_{s+1}) = u_s^2 Q(z_s^{-1},1,z_{s+1}).
	\label{eqE12b}
	\end{equation}
Using this expression in (\ref{eqE12a}) provides the final expression
	\begin{equation}
	R_s(u_s,u_{s-1},\ldots) 
	= u_s^2 \widetilde{R}_s(z_s,z_{s-1},\ldots), 
	\label{eqE12}
	\end{equation}
	\begin{equation}
	\widetilde{R}_s(z_s,z_{s-1},\ldots) = 
	\int Q(z_s^{-1},1,z_{s+1}) \widetilde{P}_{\textrm{uni}}(z_{s+1}|z_s,z_{s-1},\ldots) dz_{s+1}.
	\label{eqE12c}
	\end{equation}
The universality of $\widetilde{P}_{\textrm{uni}}$ automatically implies the universality of the reduced system function $R_s$. Furthermore, the expressions show that $R_s$ is a homogeneous function of its arguments of degree $2$, just like the original nonlinearity $Q_s$. Short-range dependence of $\widetilde{P}_{\textrm{uni}}(z_{s+1}|z_s,z_{s-1},\ldots)$ on its arguments leads to the similar property for $R_s(u_s,u_{s-1},\ldots)$: this function depends essentially on a few variables $u_s,u_{s-1},\ldots$, while the dependence on $u_n$ becomes very weak with decreasing $n$. 

We arrived to the important and rigorous conclusion that the third Kolmogorov hypothesis yields the universal law (\ref{eqE12}), (\ref{eqE12c}) describing the dynamics of the last shell $u_s$ in the system (\ref{eqE7}), (\ref{eqE8}). This deterministic dynamical system governs the evolution of the reduced probability density $P_s$ in Eq.~(\ref{eqE5}) in the developed turbulent regime.  

Note that the standard Large Eddy Simulations (LES) formulation of the Navier--Stokes equations involves modeling of the (effective) turbulent eddy viscosity \cite{meneveau2000scale, pope2001turbulent, sagaut2006large,lesieur2005large}. Such viscosity can be introduced explicitly in terms of velocity field or defined with renormalization techniques, see e.g.~\cite{zhou2010renormalization,verma2004large}. For example, it is defined in terms of the rate-of-strain tensor in the Smagorinsky model~\cite{smagorinsky1963general}. As a result, the effective viscous term  is a homogeneous function of degree 2 in the velocity field. Such an observation puts our closure for shell models in direct relation with the LES approach: 
the homogeneous function $R_s$ can be seen as a generalized term that includes the turbulent eddy viscosity. The idea of our work is to go beyond the concept of effective viscosity focused on the process of energy dissipation, by modeling a closure that describes the actual statistics at subgrid shells. As such, the proposed analysis of subgrid closures in shell models becomes a useful theoretical tool for testing optimal strategies, which may be potentially extended to the LES schemes for the NS equations. 

\section{Application to the Desnyansky--Novikov shell model}
\label{sec4b}
\begin{figure}
\centering
\includegraphics[width = 0.5\textwidth]{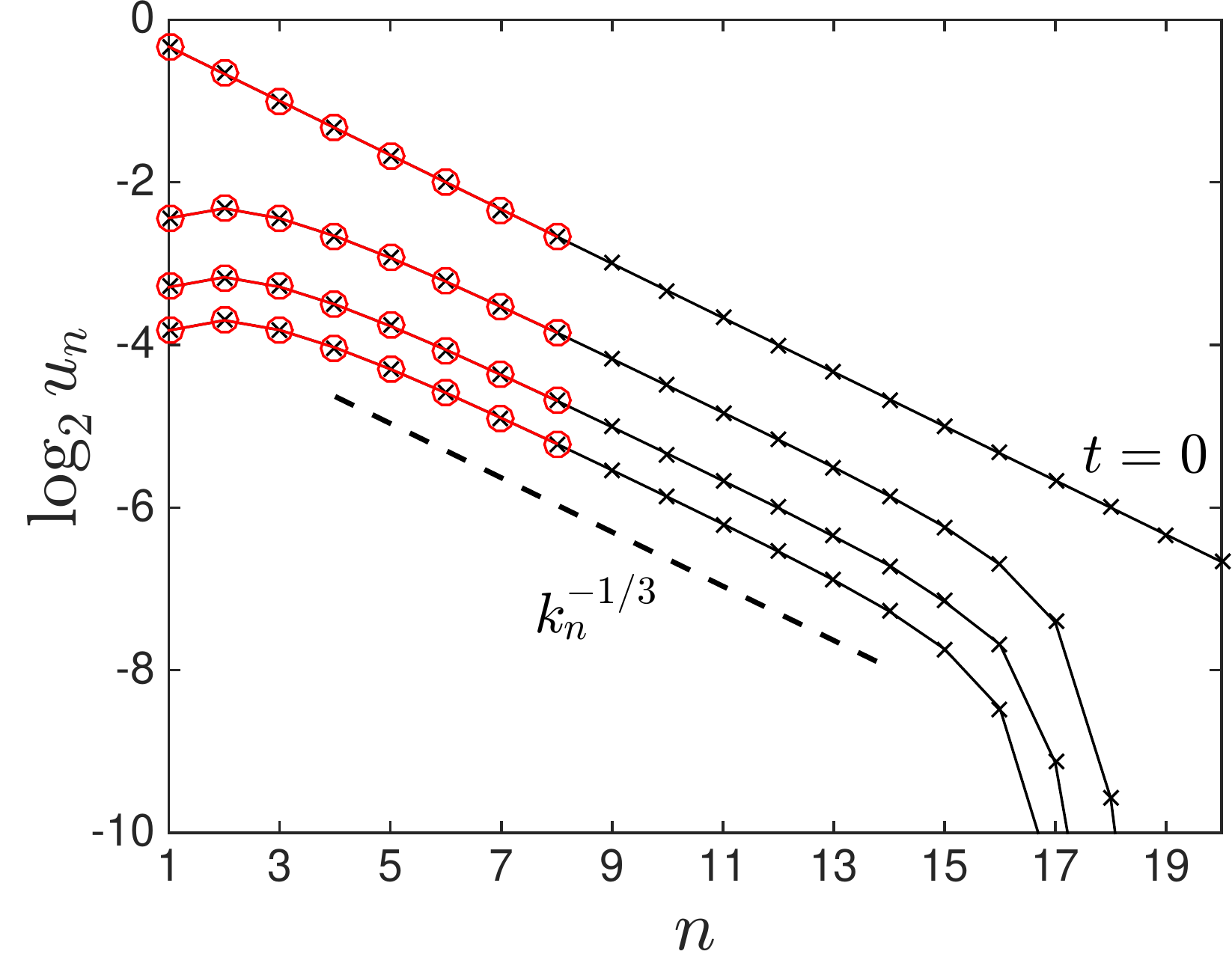}
\caption{Black lines with crosses are solutions of the full Desnyansky--Novikov model at times $t = 0,1,2,3$ (lower curves correspond to larger times). Red lines with circles show corresponding solutions for the reduced system. The dashed line marks the slope $\propto\!k_n^{-1/3}$ in the inertial interval.}
\label{fig0}
\end{figure}

Typical models with simple first-neighbor coupling as  (\ref{eqE0b}) develop a 
 non-chaotic (non-turbulent) behavior. However, we still can use such models for demonstrating basic principles of the reduction, before considering more sophisticated models in the next section. 
Let us consider the Desnyansky--Novikov model~\cite{desnyansky1974evolution,bell1977nonlinear} 
defined by the nonlinear term of the form
	\begin{equation}
	Q(u_{n-1},u_n,u_{n+1}) = u_{n-1}^2-\lambda u_n u_{n+1}.
	\label{eqF1}
	\end{equation}
The model possesses the energy $E = \frac{1}{2}\sum u_n^2$ 
as an inviscid invariant and has a non-chaotic time evolution.
Solutions of equations (\ref{eqE0}) develop a power-law tail 
	\begin{equation}
	u_n \approx ak_n^{-1/3}
	\label{eqF2}
	\end{equation}
in the inertial interval, Fig.~\ref{fig0}. Here $a > 0$ is an arbitrary factor generally depending on time, which is related to the energy flux from large to small scales. Such a tail can be interpreted as a shock wave for a continuous representation of the model~\cite{mailybaev2015continuous}, in close analogy with the Burgers equation.

Though the shell model dynamics is regular, the Kolmogorov hypothesis holds in the inertial interval. As follows from Eq.~(\ref{eqF2}), all the multipliers 
	\begin{equation}
	z_n = \frac{u_n}{u_{n-1}} = \lambda^{-1/3}
	\label{eqF3}
	\end{equation}
are constant. The corresponding universal probability density becomes the Dirac delta-function as
	\begin{equation}
	\widetilde{P}_{\mathrm{uni}}(z_{s+1}|z_s,\ldots) = \delta(z_{s+1}-\lambda^{-1/3}).
	\label{eqF4}
	\end{equation}
Relation (\ref{eqE12c}) with $Q$ from (\ref{eqF1}) yields 
	\begin{equation}
	\widetilde{R}_s
	= \int \left(\frac{1}{z_s^2}-\lambda z_{s+1}\right)\delta(z_{s+1}-\lambda^{-1/3})dz_{s+1}
	= \frac{1}{z_s^2}-\lambda^{2/3}.
	\label{eqF5}
	\end{equation}
Finally, we find the reduced system function from (\ref{eqE12}) as
	\begin{equation}
	R_s(u_s,u_{s-1}) = u_{s-1}^2-\lambda^{2/3} u_s^2.
	\label{eqF6}
	\end{equation}
The reduced system (\ref{eqE7}), (\ref{eqE8}) becomes 
	\begin{equation}
	\dot u_n = \left\{\begin{array}{ll}
	k_n\left(u_{n-1}^2-\lambda u_n u_{n+1}\right),& n = 1,\ldots,s-1;
	\\[5pt]
	k_s\left(u_{s-1}^2-\lambda^{2/3} u_s^2\right), & n = s.
	\end{array}\right.
	\label{eqF7}
	\end{equation}

For a numerical test, we consider the Desnyansky--Novikov model with $n = 20$ shells, viscosity $\nu = 10^{-7}$ and wavenumbers $k_n = \lambda^n$ with $\lambda = 2$. As a boundary condition, we take $u_0 = 0$ and consider a decaying solution from the initial data $u_n(0) = k_n^{-1/3}$. The solution is shown by black lines with crosses in Fig.~\ref{fig0}. It has viscous range around the shell $17$ and the power-law dependence (\ref{eqF2}) in the inertial interval $4 \lesssim n \lesssim 15$. The reduced system (\ref{eqF7}) was integrated with $s = 8$ shells, and the numerical results are presented in Fig.~\ref{fig0} by red circles demonstrating an excellent match with the full model solution.

\section{Application to the Sabra shell model}
\label{sec6}

It is straightforward to extend the results of Sections~\ref{sec3} and \ref{sec4} to more general shell models, where shell variables are complex numbers as in the GOY or Sabra shell models~\cite{gledzer1973system,ohkitani1989temporal,l1998improved} or defined in term  of sets of variables (vectors) \cite{eggers1991does,benzi_1996_Helical_shell_models,de2015inverse,lessinnes2009energy}. Equations of motion for such models have the same structure (\ref{eqE0}), but the nonlinear term $Q_n$ may depend on several shells from each side.

In this section, we formulate the reduction for the Sabra shell model~\cite{l1998improved}, which is characterized by complex shell variables $u_n \in \mathbb{C}$. The nonlinear term $Q_n$ in (\ref{eqE0}) describes the interaction with two neighbors given by 
	\begin{equation}
	Q_n = Q(u_{n-2},\ldots,u_{n+2}) = 
	i\left(\lambda u_{n+2}u_{n+1}^*+bu_{n+1}u_{n-1}^*
	-c\lambda^{-1}u_{n-1}u_{n-2}\right),
	\label{eqE19}
	\end{equation}
where the shells $u_0$ and $u_{-1}$ must be specified by boundary conditions. The choice $c = -1/\lambda$ and $b = -1-c$ corresponds to the so-called 3D regime and leads to the two inviscid invariants: the energy $E = \frac{1}{2}\sum |u_n|^2$ and the helicity $H = \frac{1}{2}\sum (-1)^nk_n|u_n|^2$.

The probability density of this system evolves under the continuity equation written similarly to (\ref{eqE1}) as
	\begin{equation}
	\frac{\partial P}{\partial t} 
	+ \sum_{n = 1}^N 
	\left[
	\frac{\partial}{\partial a_n} \left(\dot a_n P\right)
	+\frac{\partial}{\partial b_n} \left(\dot b_n P\right)
	\right]
	 = 0,
	\label{eqE15}
	\end{equation}
where we denoted $u_n = a_n+ib_n$.
The analogous derivation as in Section~\ref{sec3} yields the description for the reduced probability distribution $P_s(u_1,\ldots,u_s)$ in the form
	\begin{equation}
	\frac{\partial P_s}{\partial t} 
	+ \sum_{n = 1}^N 
	\left[
	\frac{\partial}{\partial a_n} \left(k_n A_n P_s\right)
	+\frac{\partial}{\partial b_n} \left(k_n B_n P_s\right)
	\right]
	 = 0,
	\label{eqE16}
	\end{equation}
where $s$ is a fixed shell number from the inertial interval, and the functions $A_n+iB_n = R_n(u_1,\ldots,u_s)$ are defined below. For complex variables $u_n$, this equation corresponds to the evolution of probability density for the reduced dynamical system 
	\begin{equation}
	\dot u_n = k_nR_n,\quad n = 1,\ldots,s.
	\label{eqE7x}
	\end{equation}

Similarly to Eqs.~(\ref{eqE8}) and (\ref{eqE9}) one derives
	\begin{equation}
	R_n = \left\{\begin{array}{ll}
	Q_n, & n = 1,\ldots,s-2;\\[7pt]
	\displaystyle
	\langle Q_n | u_s, u_{s-1}\ldots \rangle,& n = s-1,s,
	\end{array}\right.
	\label{eqE17}
	\end{equation}
where the first $s-2$ equations of the shell model remain unchanged. The last two equations are given by the conditional averages 
	\begin{equation}
	\langle Q_n | u_s, u_{s-1}\ldots \rangle 
	= \int Q_n P_{\textrm{cond}}(u_{s+2},u_{s+1}|u_s,\ldots,u_1;t) 
	\prod_{m = s+1}^{s+2} da_mdb_m.
	\label{eqE18}
	\end{equation}
Again, our construction shows that the subgrid scheme is given by a deterministic system of equations.
The time-independence and universality of this conditional average follows from the Kolmogorov hypothesis formulated for multipliers as we demonstrate below. 

Let us introduce the complex multipliers $z_n \in \mathbb{C}$ as~\cite{benzi1993intermittency,eyink2003gibbsian} 
	\begin{equation}
	z_n 
	= w_{n}e^{i\Delta_n} 
	,\quad
	w_n = \left|\frac{u_{n}}{u_{n-1}}\right|,\quad
	\Delta_n = \arg u_n-\arg u_{n-1}-\arg u_{n-2}.
	\label{eqSM3}
	\end{equation}
Here the phases $\Delta_n$ are chosen to  be invariant under the phase symmetry
	\begin{equation}
	u_n \mapsto u_n e^{i\theta_n},\quad \theta_n = \theta_{n-1}+\theta_{n-2}, \quad 
	n = 1,2,\ldots
	\label{eqE21}
	\end{equation}
which is an analog in the Sabra model of the 
 physical space homogeneity \cite{l1998improved}. 
The combination of phases given by $\Delta_n$ is important because it is strictly connected to the existence of a mean forward energy cascade 
(see expression (\ref{eq1d}) and discussion thereof).  
It is easy to see that there is one-to-one correspondence between the multipliers $z_1,\ldots,z_s$ and the shell variables $u_1,\ldots,u_s$ with $u_0$ and $u_{-1}$ given by boundary conditions (except for a zero-measure subset when some $u_n = 0$). Thus, an argument similar to the one used in Eqs.~(\ref{eqE10}) and (\ref{eqE10b}) can be applied to the conditional probability $P_{\textrm{cond}}$ in (\ref{eqE18}). Namely, one can use the third Kolmogorov hypothesis for expressing this function as
	\begin{equation}
	P_{\textrm{cond}}(u_{s+2},u_{s+1}|u_s,\ldots,u_1;t) \prod_{m = s+1}^{s+2} da_mdb_m
	= \widetilde{P}_{\textrm{uni}}(z_{s+2},z_{s+1}|z_s,\ldots) \prod_{m = s+1}^{s+2} dx_mdy_m,
	\label{eqE18b}
	\end{equation}
where we denoted $z_n = x_n+iy_n$ and $\widetilde{P}_{\textrm{uni}}$ is a universal function describing the conditional distribution for multipliers in the developed turbulent regime. The function $\widetilde{P}_{\textrm{uni}}$ is expected to have a short-range dependence on its arguments, i.e., the dependence on $z_n$ for $n \le s$ gets very weak with decreasing $n$. The universality property was thoroughly studied numerically in \cite{benzi1993intermittency,eyink2003gibbsian}. 

According to (\ref{eqE17}) the reduced system (\ref{eqE7x}) contains the two unknown functions on the right-hand sides, $R_{s-1}(u_s,u_{s-1},\ldots)$ and $R_s(u_s,u_{s-1},\ldots)$. Using the explicit form (\ref{eqE19}) of the nonlinear term in the conditional average (\ref{eqE18}) for $n = s-1$ yields
	\begin{equation}
	\begin{array}{rcl}
	{R}_{s-1} 
	& = & \langle Q_{s-1} | u_s, u_{s-1}\ldots \rangle \\[5pt]
	& = & i\left(\lambda u_{s}^* \langle u_{s+1}| u_s, u_{s-1}\ldots \rangle
	+bu_{s}u_{s-2}^*
	-c\lambda^{-1}u_{s-2}u_{s-3}\right)\\[5pt]
	& = & i\left(\lambda |u_{s}|^2e^{i\arg u_{s-1}} \langle z_{s+1}| z_s, z_{s-1}\ldots \rangle
	+bu_{s}u_{s-2}^*
	-c\lambda^{-1}u_{s-2}u_{s-3}\right),
	\end{array}
	\label{eqTN5}
	\end{equation}
where we used (\ref{eqSM3}) and the fact that the conditional average for given $u_s,u_{s-1},\ldots$ is equal to the conditional average for given $z_s,z_{s-1},\ldots$. Similar computation for $R_{s}$ yields
	\begin{equation}
	\begin{array}{rcl}
	{R}_{s} 
	& = & \langle Q_{s} | u_s, u_{s-1}\ldots \rangle \\[5pt]
	&=& \displaystyle
	i\left(\lambda \langle u_{s+2}u_{s+1}^*| u_s, u_{s-1}\ldots\rangle
	+bu_{s-1}^*\langle u_{s+1}| u_s, u_{s-1}\ldots\rangle
	-c\lambda^{-1}u_{s-1}u_{s-2}\right) \\[5pt]
	&=& \displaystyle
	i\left(\lambda |u_s| u_{s}
	\left\langle |z_{s+1}|^2z_{s+2}\big| z_s, z_{s-1}\ldots\right\rangle
	+b|u_{s-1}|u_s\langle z_{s+1}| z_s, z_{s-1}\ldots\rangle
	-c\lambda^{-1}u_{s-1}u_{s-2}\right).
	\end{array}
	\label{eqTN6}
	\end{equation}
Combining (\ref{eqTN5}) and (\ref{eqTN6}), we write the final expressions for the reduced system as
	\begin{equation}
	\begin{array}{rcl}
	{R}_{s-1}(u_s,u_{s-1},\ldots) 
	& = & i\left(\lambda |u_{s}|^2 e^{i\arg u_{s-1}} \mathcal{A}_1
	+bu_{s}u_{s-2}^*
	-c\lambda^{-1}u_{s-2}u_{s-3}\right),\\[5pt]
	{R}_{s}(u_s,u_{s-1},\ldots) 
	&=& \displaystyle
	i\left(\lambda 
	|u_s|u_{s}\mathcal{A}_2
	+b |u_{s-1}|u_s\mathcal{A}_1
	-c\lambda^{-1}u_{s-1}u_{s-2}\right).
	\end{array}
	\label{eqTN9}
	\end{equation}
where	
	\begin{equation}
	\mathcal{A}_1 = 
	\langle z_{s+1}| z_s, z_{s-1}\ldots \rangle
	,\quad 
	\mathcal{A}_2 = 
	\left\langle |z_{s+1}|^2z_{s+2}\big| z_s, z_{s-1}\ldots
	\right\rangle
	.
	\label{eqTN8}
	\end{equation}
	
By construction, $\mathcal{A}_1$ and $\mathcal{A}_2$ 
are functions of $z_s,z_{s-1},\ldots$ or, equivalently, homogeneous functions of zero degree with respect to shell variables $u_s,u_{s-1},\ldots$. As a result, the functions $R_{s-1}$ and $R_s$ in (\ref{eqTN9}) are homogeneous of degree 2 with respect to shell variables $u_n$. 
Note that the multipliers in our approach are used for justification of the universality, but the final expressions can be written in terms of the original shell variables $u_n$. 

\subsection{Simple subgrid models}
\label{sec:num1}

\begin{figure}
\centering
\includegraphics[width = 0.75\textwidth]{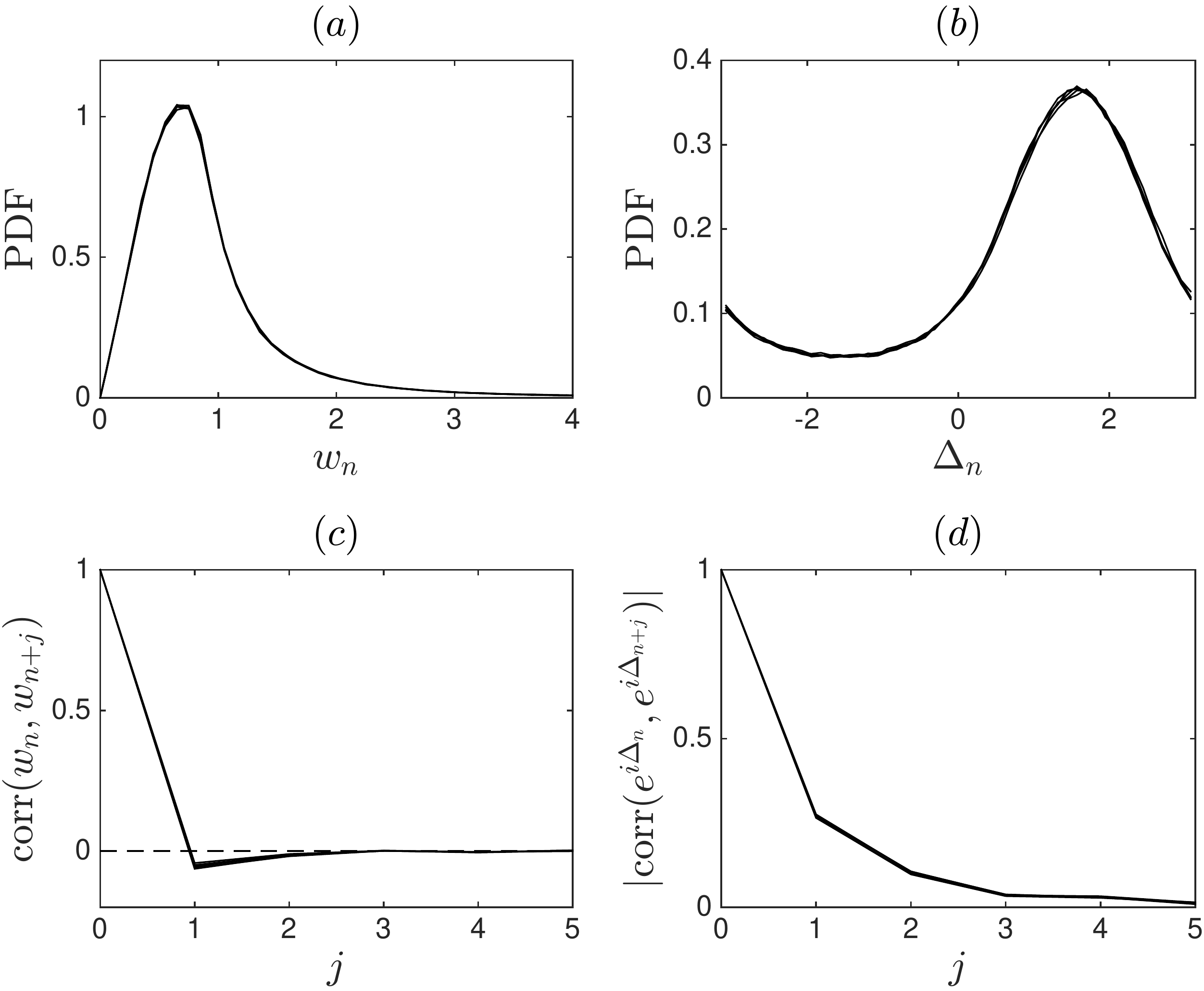}
\caption{Probability density functions of (a) speed multipliers $w_n$ and (b) phases $\Delta_n$. (c,d) Corresponding correlation coefficients as functions of shell separation $j$. Each panel shows five curves corresponding to $n = 13,\ldots,17$. These curves collapse to a scale-invariant universal distribution.}
\label{fig1}
\end{figure}

For computing the statistics of multipliers, we performed a single long-time simulation with the inter-shell ratio $\lambda = 2$ and the viscosity $\nu = 10^{-12}$ (the viscous range starts around $n_K\approx 30$) and constant boundary conditions with $u_0 = u_{-1}/2 = 1+i$. All the results in this section are obtained using $10^6$ samples $(u_1,\ldots,u_N)$ equally spaced in time. Fig.~\ref{fig1} (a,b) shows the PDFs for the absolute values $w_n = |z_n|$ and phases $\Delta_n = \arg z_n$ for different $n$, confirming the scale-invariance of the distribution for multipliers in agreement with earlier results~\cite{benzi1993intermittency,eyink2003gibbsian}.  Fig.~\ref{fig1} (c,d) demonstrates the corresponding correlation functions confirming the short-range property, i.e., a rapid decay of correlations with the shell separation. The correlation is larger for phases than for absolute values. The main consequence is that, with such rapidly decaying correlations, one may hope to obtain a reasonably accurate approximate subgrid model by keeping very few multipliers in the conditional averages (\ref{eqTN8}). 
 
In this section, we introduce three low-order approximations for the functions (\ref{eqTN8}) that uniquely determine the corresponding subgrid models given by (\ref{eqE7x}), (\ref{eqE17}) and (\ref{eqTN9}). The first model will be denoted by $\mathrm{SM_K}$ and called the Kolmogorov closure. It is given by the probability function
	\begin{equation}
	\widetilde P_{uni}(z_{s+2},z_{s+1}) = \delta(z_{s+2}-i\lambda^{-1/3})\,\delta(z_{s+1}-i\lambda^{-1/3}).
	\label{eqPF1}
	\end{equation}
Here the product of Dirac delta-functions simply means that both multipliers have deterministic absolute values $w_{s+2} = w_{s+1} = \lambda^{-1/3}$ according to the Kolmogorov scaling law, and their phases $\Delta_{s+2} = \Delta_{s+1} = \pi/2$ are fixed at the most probable values, see Fig.~\ref{fig1}(b). In this case, expressions (\ref{eqTN8}) yield
	\begin{equation}
	\mathrm{SM_K}:\quad
	\mathcal{A}_1 = 
	i\lambda^{-1/3} \approx 0.79i,\quad 
	\mathcal{A}_2 = 
	i\lambda^{-1} = 0.5i.
	\label{eqPF2}
	\end{equation}
The next model, denoted by $\mathrm{SM_0}$, is considered to be a zero-order approximation based on numerical information obtained from the unclosed original equations. Namely, we take the  values
	\begin{equation}
	\mathrm{SM_0}:\quad
	\mathcal{A}_1 =  \langle z_{s+1}\rangle \approx 0.40i,\quad 
	\mathcal{A}_2 = \langle |z_{s+1}|^2z_{s+2}\,\big|\, z_s = i\lambda^{-1/3} \rangle \approx
	0.37i,
	\label{eqPF3}
	\end{equation}
by supposing to consider the minimal degree of correlation in  (\ref{eqTN8}).
In particular, the value for $\mathcal{A}_1$ has been estimated from an unconditional average of the full viscous 
 unclosed model, while for $\mathcal{A}_2$ we imposed the minimal constraints
by fixing the first resolved multiplier to its Kolmogorov value. The conditioning in the latter expression is necessary, because the unconditional average $\langle |z_{s+1}|^2z_{s+2}\rangle$ would diverge. The divergence is related to shell variables $u_s$ passing close to the origin, which leads to large $|z_{s+1}| = |u_{s+1}/u_s|$ simultaneously with small $|z_{s}| = |u_{s}/u_{s-1}|$, see~\cite{eyink2003gibbsian}.  Thus, this defect can be avoided by excluding the events with  small $z_s$. Note that the numerical values in (\ref{eqPF3}) are quite different from the Kolmogorov prediction in (\ref{eqPF2}), caused by cancellations due to a large spread of the phases.

\begin{figure}
\centering
\includegraphics[width = 0.9\textwidth]{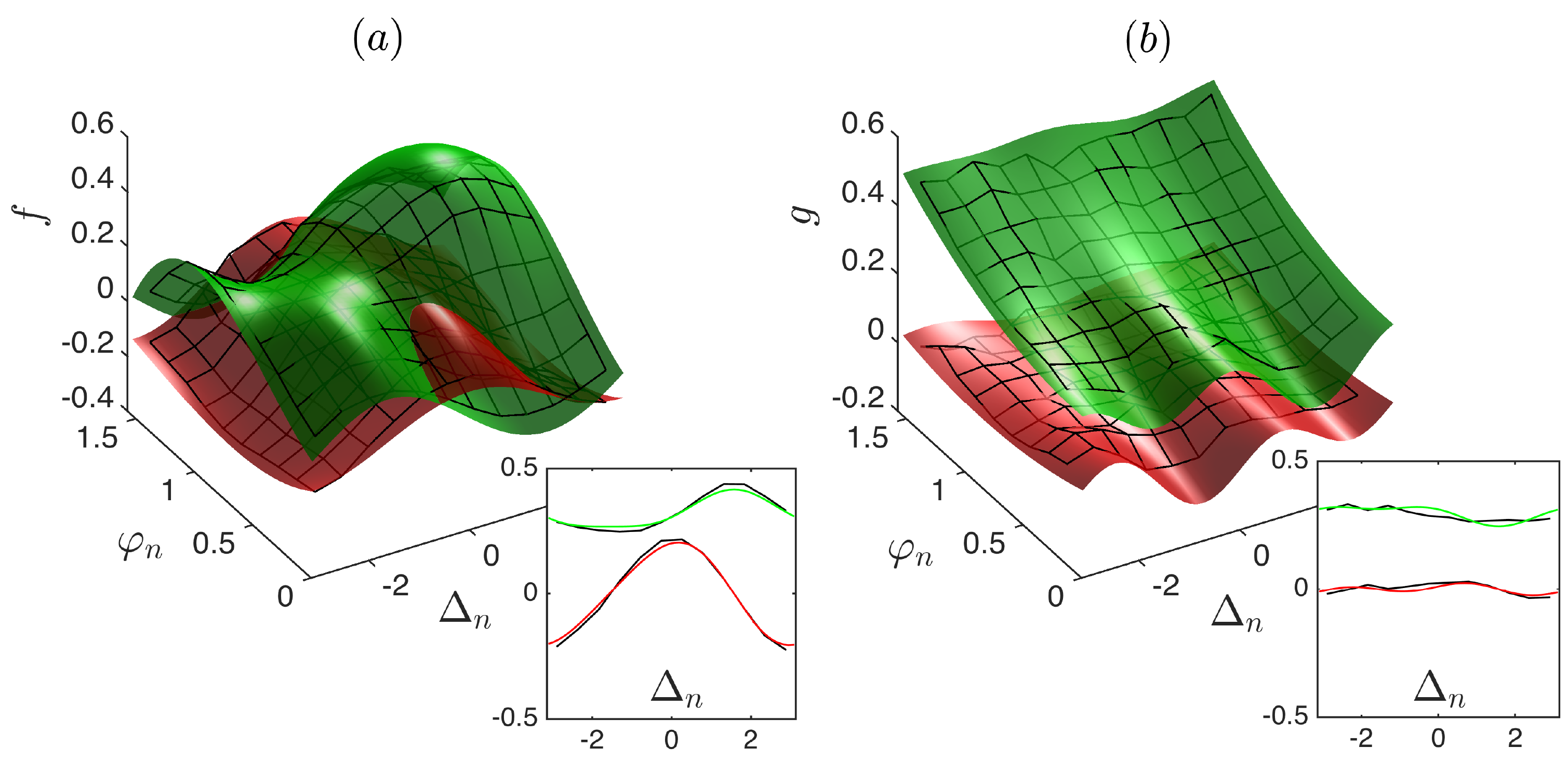}
\caption{Functions (a) $f(\varphi_n,\Delta_n)$ and (b) $g(\varphi_n,\Delta_n)$ of the subgrid model $\mathrm{SM_1}$. Red and green surfaces correspond to real and imaginary parts, respectively. Black grids indicate the values obtained by averaging with the numerical simulation results. {The insets demonstrate the comparison for cross-sections with $\varphi_n = \pi/4$.} }
\label{figB1}
\end{figure}

Finally, we consider the model denoted by $\mathrm{SM_1}$, which is obtained as a first-order approximation by considering the averages conditioned to the single multiplier $z_s = |u_s/u_{s-1}|e^{i\Delta_s}$ in (\ref{eqTN8}), i.e.,
	\begin{equation}
	\mathrm{SM_1}:\quad
	\mathcal{A}_1(z_s) =  \langle z_{s+1}|z_s\rangle,\quad 
	\mathcal{A}_2(z_s) = \langle |z_{s+1}|^2z_{s+2}\big| z_s \rangle.
	\label{eqPF4}
	\end{equation}
The averages are  determined numerically from 
the results of a direct numerical simulation of the full unclosed equations 
as mentioned above. Due to the ``defect'' of the averages at the origin, it is convenient to use the ansatz  
	\begin{equation}
	\mathrm{SM_1}:\quad
	\mathcal{A}_1(z_s) =  \varphi_s^{-1}f(\varphi_s,\Delta_s),\quad 
	\mathcal{A}_2(z_s) = \varphi_s^{-1}g(\varphi_s,\Delta_s),\quad
	\varphi_s = \mathrm{atan}\, w_s.
	\label{eqPF5}
	\end{equation}
Here, the variable $0 \le \varphi_s < \pi/2$ is used for representing the infinite semi-interval $w_s \ge 0$. The functions $f(\varphi_s,\Delta_s)$ and $g(\varphi_s,\Delta_s)$ are universal in the inertial interval. {We compute them by fitting the simulation data with the Fourier expansion in $\Delta_s$, where coefficients are polynomial functions of $\varphi_s$, see Fig.~\ref{figB1}. The averaged data are obtained by taking mean values of the simulation results in every grid cell $(\delta\Delta_s,\delta\varphi_s)$ in Fig.~\ref{figB1}.} This yields
	\begin{equation}
	f(\varphi_s,\Delta_s) = if_0+f_1e^{i\Delta_s}+f_{-1}e^{-i\Delta_s}+if_2e^{2i\Delta_s},\quad 
	g(\varphi_s,\Delta_s) = ig_0+g_1e^{i\Delta_s}+g_{-1}e^{-i\Delta_s}+ig_{-2}e^{-2i\Delta_s},
	\label{eqPF6}
	\end{equation}
with 
	\begin{equation}
	\begin{array}{c}
	f_0 = -0.32 \varphi_s^3+ 0.23\varphi_s^2+0.43\varphi_s,\quad 
	f_1= -0.04\varphi_s^3-0.04\varphi_s^2+   0.27\varphi_s   -0.03,\\
	f_{-1} = 0.13\varphi_s^3   -0.32\varphi_s^2+    0.1\varphi_s+    0.12,\quad
	f_{2} = -0.13 \varphi_s^3+   0.34\varphi_s^2   -0.21\varphi_s,\\
	g_0 = 0.07\varphi_s^2+    0.08\varphi_s+    0.19,\quad
	g_1 = -0.03\varphi_s^2+    0.06\varphi_s   -0.04,\\
	g_{-1} = -0.04\varphi_s^2+    0.06\varphi_s,\quad
	g_{-2} = 0.04\varphi_s^2   -0.10\varphi_s+    0.07.
	\end{array}
	\label{eqPF7}
	\end{equation}
In these approximations, we kept the Fourier modes for $\Delta_s$, which had the dominant contribution, while the coefficients depending on $\varphi_s$ were approximated with low-order polynomials neglecting the coefficients smaller than $0.02$. All functions in (\ref{eqPF7}) appear to be real due to the inherent symmetries of the Sabra model.
Fig.~\ref{figB1} demonstrated the comparison of expressions (\ref{eqPF6}) with the same functions found numerically by averaging expressions (\ref{eqPF4}) in the inertial interval. 

\subsection{Numerical tests}
\label{sec:numtest1}
 
The numerical tests are carried out for the three subgrid models from the previous section with $s = 15$ and $20$ shells. We do this in the time interval $0 \le t \le 10^3$ with the constant boundary conditions 
	\begin{equation}
	u_0 = 2,\quad u_{-1} = 1.
	\label{eqNT1}
	\end{equation}
For initial conditions we take the Kolmogorov state $u_n = k_n^{-1/3}e^{i\phi_n}$ with random phases $\phi_n$.
The comparison is also made with the simulation of the full viscous model (\ref{eqE0}), (\ref{eqE19}) with total $N = 40$ shells and viscosity $\nu = 10^{-12}$ (the viscous range starts around shell $n_K \sim 30$). 

\begin{figure}
\centering
\includegraphics[width = 0.95\textwidth]{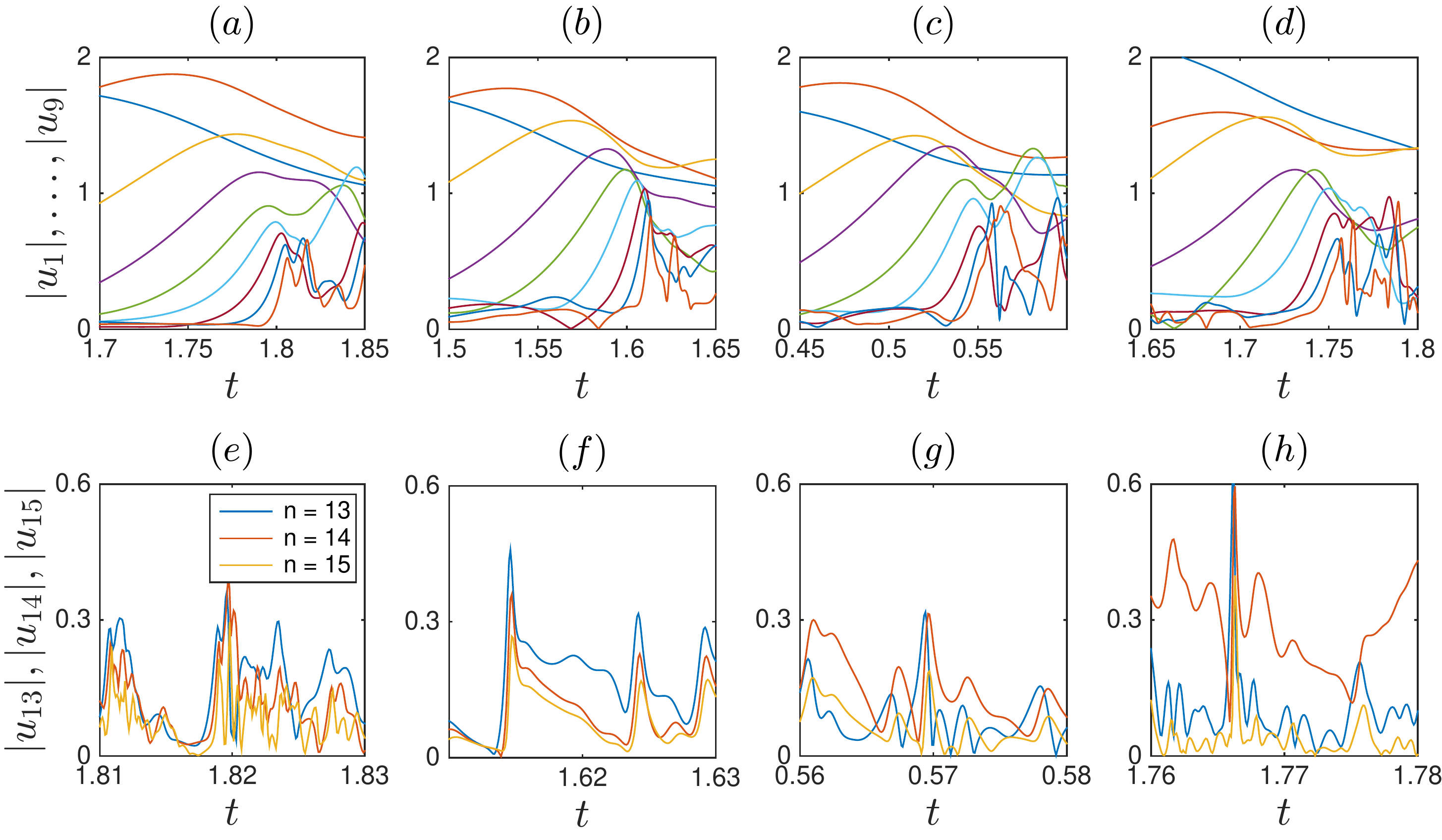}
\caption{Qualitative comparison of the dynamics of absolute values for large-scale shell variables $u_1,\ldots,u_{9}$ (first row, curves from top to bottom) and for the 
cutoff variables $u_{s-2},u_{s-1},u_{s}$ (second row) with $s = 15$. (a, e) The full model with viscosity $\nu = 10^{-12}$. The subgrid models: (b, f) SM${}_K$, (c, g) SM${}_0$, (d, h) SM${}_1$. Time windows are different for different panels. 
}
\label{figN2}
\end{figure}

Simulation results for some representative interval of time are compared in Fig.~\ref{figN2}. One can see that the dynamics at large scales (first row) is qualitatively similar for all models, but small scales (second row) demonstrate some qualitative differences. In particular, we notice a tendency to lock among the last three variables much more pronounced than in the unclosed case, 
an indication that phase correlation is probably not fully correct. Note that obviously we do not expect a detailed correspondence of the solutions, since subgrid models are designed to describe a probability distribution rather than a particular solution. 

\begin{figure}
\centering
\includegraphics[width = 0.67\textwidth]{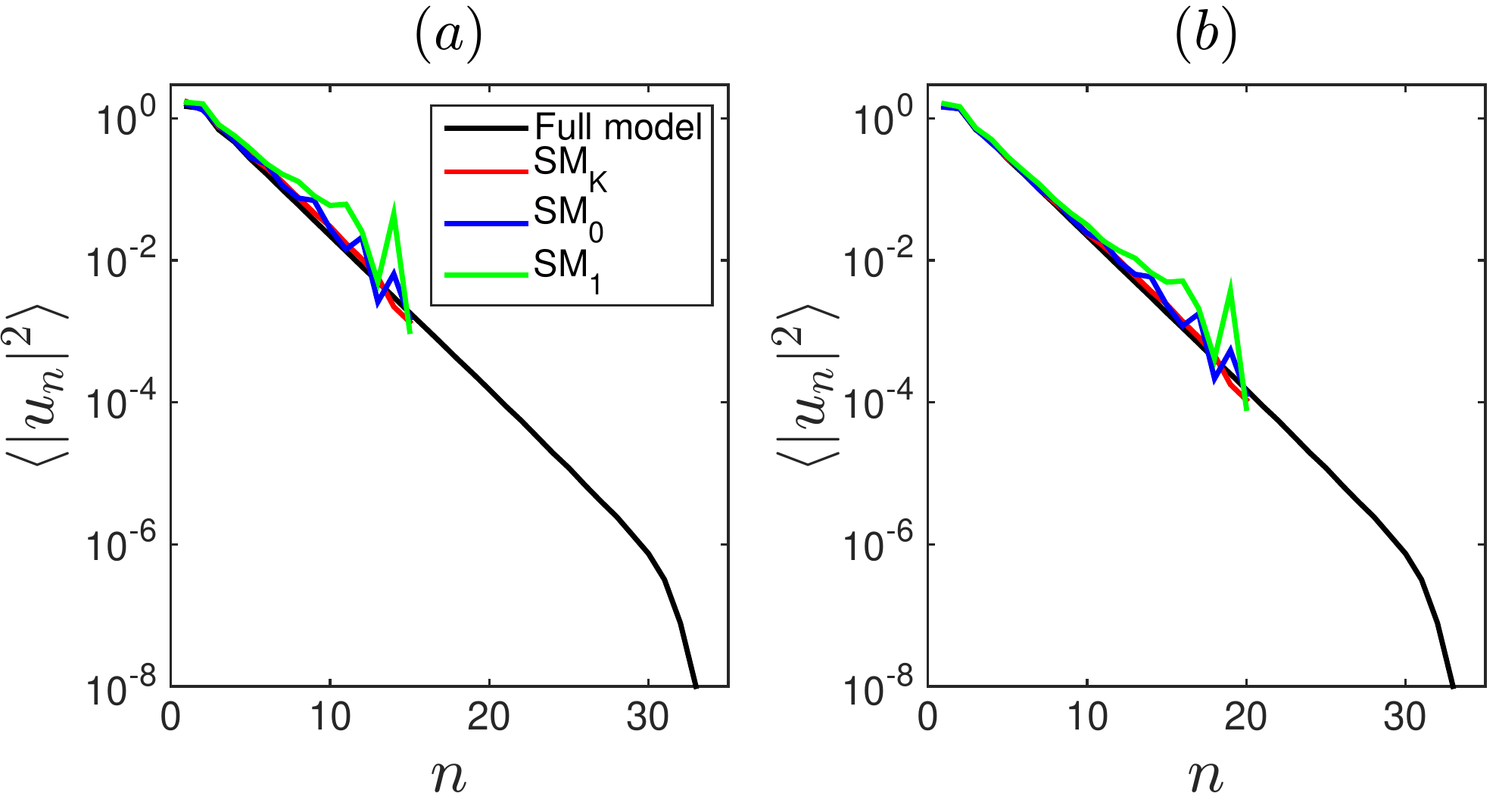}
\caption{Energy spectrum for the full model (black) is compared with different subgrid models: $\mathrm{SM_K}$ (red), $\mathrm{SM_0}$ (blue) and $\mathrm{SM_1}$ (green). The results are shown for the subgrid models with cutoff at
(a) $s = 15$ shells and (b) $s = 20$ shells. }
\label{figEn}
\end{figure}

\begin{figure}
\centering
\includegraphics[width = 1\textwidth]{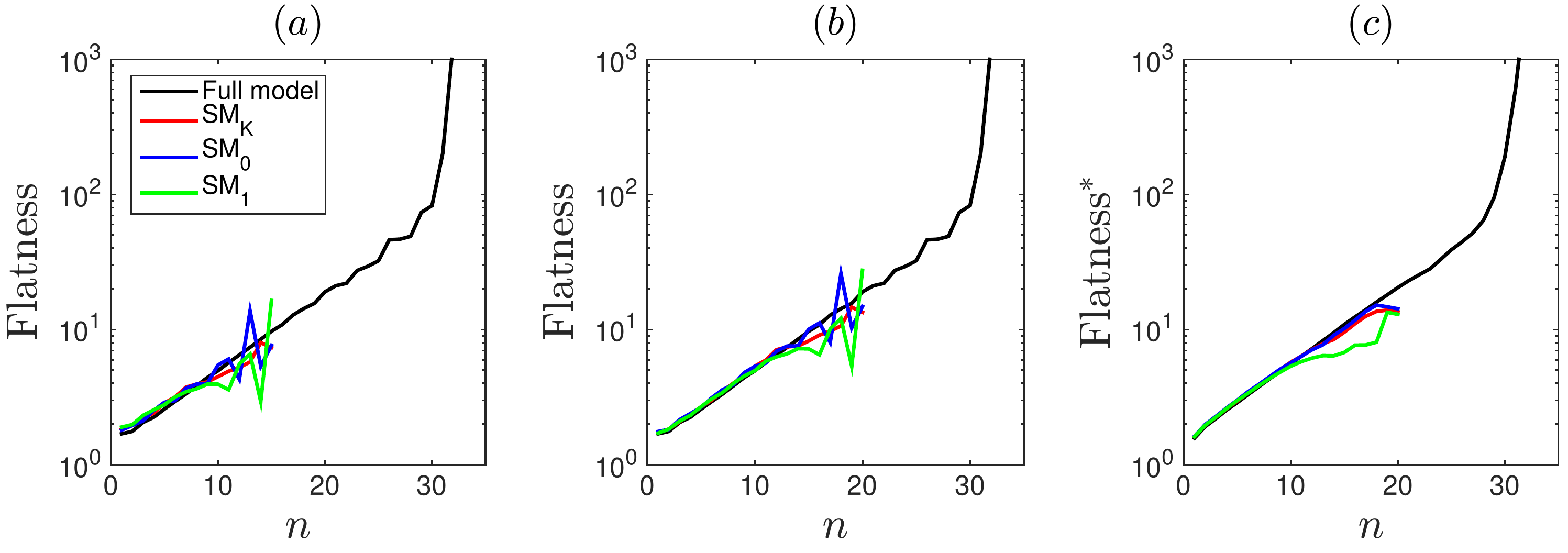}
\caption{Flatness, $F_n$, for the full and subgrid models with (a) $s = 15$ and (b) $s = 20$. (c) The flatness computed in terms of energy flux, $F_n^\Pi$, for $s = 20$.
}
\label{figN6}
\end{figure}

The differences among the models can be seen more clearly and systematically in Fig.~\ref{figEn} presenting the time-averaged energy spectra:
$$
E_n = \langle |u_n|^2 \rangle.
$$
 One can see that, indeed, large deviations are observed in the region of large shell numbers $n \sim s$ (small scales near cutoff), while good agreement is attained at smaller $n$ (larger scales). Surprisingly, more elaborated subgrid models do not show improved results, with a better match given by the simplest SM${}_K$ model. We postpone a discussion on why this happens to the next section, and concentrate on describing different aspects of solutions for different models now. From Fig.~\ref{figEn} one can see that the deviations demonstrated by every model do not depend on the total number of the shells $s$: they repeat the same pattern for $n \sim s$ that converges to the full model results for smaller $n$. Similar results are observed for other velocity moments as well. In particular, Fig.~\ref{figN6} presents the results for the flatness: 
$$
F_n= \frac{\langle |u_n|^4 \rangle}{\langle |u_n|^2 \rangle^2}
$$
 demonstrating similar type of discrepancies at final shell numbers.

\begin{figure}
\centering
\includegraphics[width = 1\textwidth]{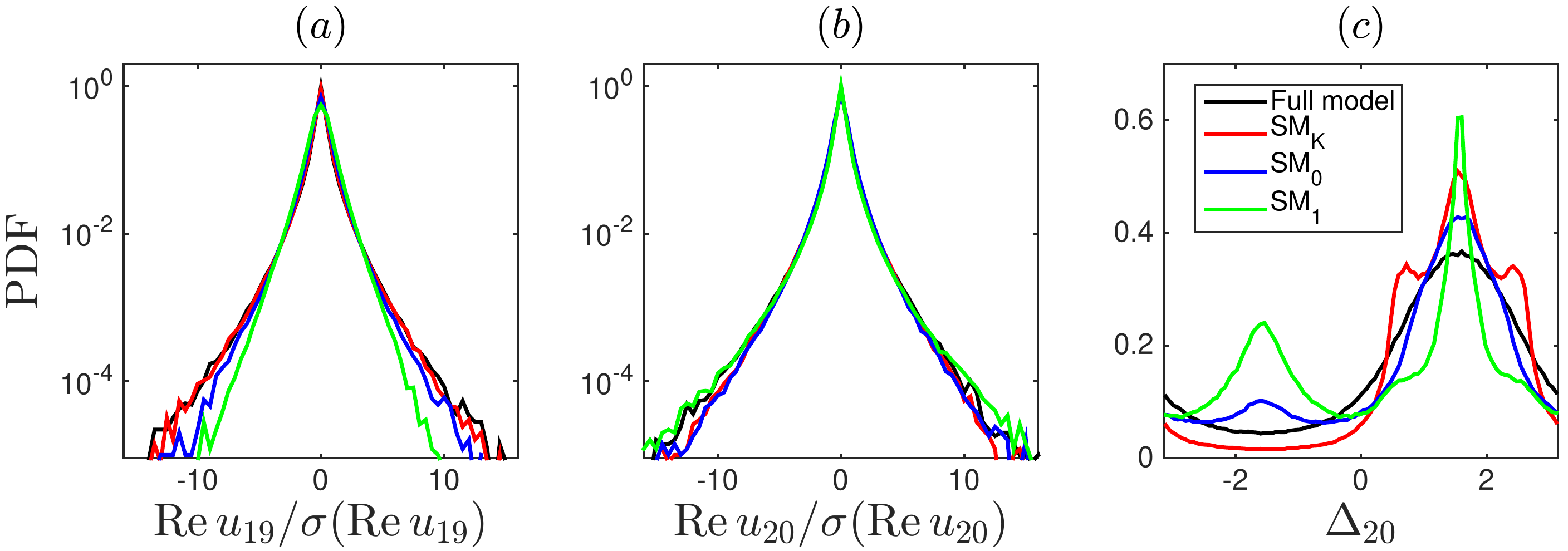}
\caption{PDFs of shell variables for the full and subgrid models with $s = 20$.  Real parts of shell variables (a) $u_{19}$ and (b) $u_{20}$; PDFs are normalized with standard deviation. (c) PDFs for the phase variable $\Delta_{20}$.
}
\label{figB5}
\end{figure}

Though showing rather large deviations for average values, the subgrid models describe probability of non-Gaussian rare events (intermittency) reasonably well. Fig.~\ref{figB5}(a,b) shows the normalized PDFs in the logarithmic vertical scale, where fat tails are well reproduced. Still, also in this case, the models SM${}_1$ or SM${}_0$ are less accurate than a simple model SM${}_K$ based on the Kolmogorov closure. Fig.~\ref{figB5}(c) provides the comparison of PDFs for the phase variable $\Delta_s$ at the last subgrid shell number $s = 20$, demonstrating a considerable variation among different models. Note that the phases $\Delta_n$ determine the direction of energy flux, see Eq.~(\ref{eq1d}) below.

The next test is related to the energy flux. Recall that the energy $E = \frac{1}{2}\sum |u_n|^2$ is an inviscid invariant for the Sabra model. A general expression for the  energy flux across shell $n$ 
is given by $ \Pi_n^E = \mathrm{Im}\,(k_{n+1}u_{n+2}u_{n+1}^*u_n^*-ck_nu_{n+1}u_n^*u_{n-1}^*)$, see e.g.~\cite{biferale2003shell,ditlevsen2010turbulence,PhysRevE.91.043014}. With the phases $\Delta_n$ from (\ref{eqSM3}), this expression is written as
\begin{equation}
 \Pi_n^E 
= k_{n+1}|u_{n+2}u_{n+1}u_n|\sin\Delta_{n+2}-ck_{n}|u_{n+1}u_nu_{n-1}|\sin\Delta_{n+1}.
\label{eq1d}
\end{equation}
Expression (\ref{eq1d}) determines the non-linear contributions to the total change of the energy in the shells up to $n$.
Thus, for inviscid dynamics, the energy balance takes the form
\begin{equation}
\frac{d}{dt}\left(\frac{1}{2}\sum_{j=1}^n |u_j|^2\right) = 
\Pi_0^E-\Pi_n^E,
\label{eqExtra1}
\end{equation}
where $\Pi_0^E$ is the work per unit time done by the boundary due to nonzero values of the shells $u_{0}$ and $u_{-1}$. For subgrid models, expression for the energy flux is different for the modified shells $n = s-1$ and $s$. Direct computations for these fluxes using (\ref{eqExtra1}) and (\ref{eqE7x}), (\ref{eqE17}), (\ref{eqTN9}) yields
\begin{equation}
\Pi_{s-1}^E = k_{s-1}\,\mathrm{Im}\left[\lambda |u_s|^2|u_{s-1}|\mathcal{A}_1-cu_{s}u_{s-1}^*u_{s-2}^*\right],
\label{eqExtra2}
\end{equation}
\begin{equation}
\Pi_s^E = k_{s}\,\mathrm{Im}\left[\lambda |u_s|^3\mathcal{A}_2
-c|u_s|^2|u_{s-1}|\mathcal{A}_1\right].
\label{eqExtra3}
\end{equation}
In particular, for the SM${}_K$ model given by (\ref{eqPF2}) with $\lambda = 2$ and $c = -1/2$, one obtains the strictly positive flux at last shell as
\begin{equation}
\mathrm{SM}_K:\quad
\Pi_s^E = k_{s}|u_s|^2\left(|u_s|
+2^{-4/3}|u_{s-1}|\right).
\label{eqExtra2b}
\end{equation}
Similarly, the inviscid invariant called helicity is introduced as $H = \frac{1}{2}\sum (-1)^nk_n|u_n|^2$. 
In this case the helicity flux of the inviscid model is given by the expression
\begin{equation}
\Pi_n^H = (-1)^nk_n^2\,\mathrm{Im}\left[\lambda u_{n+2}u_{n+1}^*u_n^*
+(b-\lambda^{-1})u_{n+1}u_n^*u_{n-1}^*\right].
\label{eq1dH}
\end{equation}

\begin{figure}
\centering
\includegraphics[width = 1\textwidth]{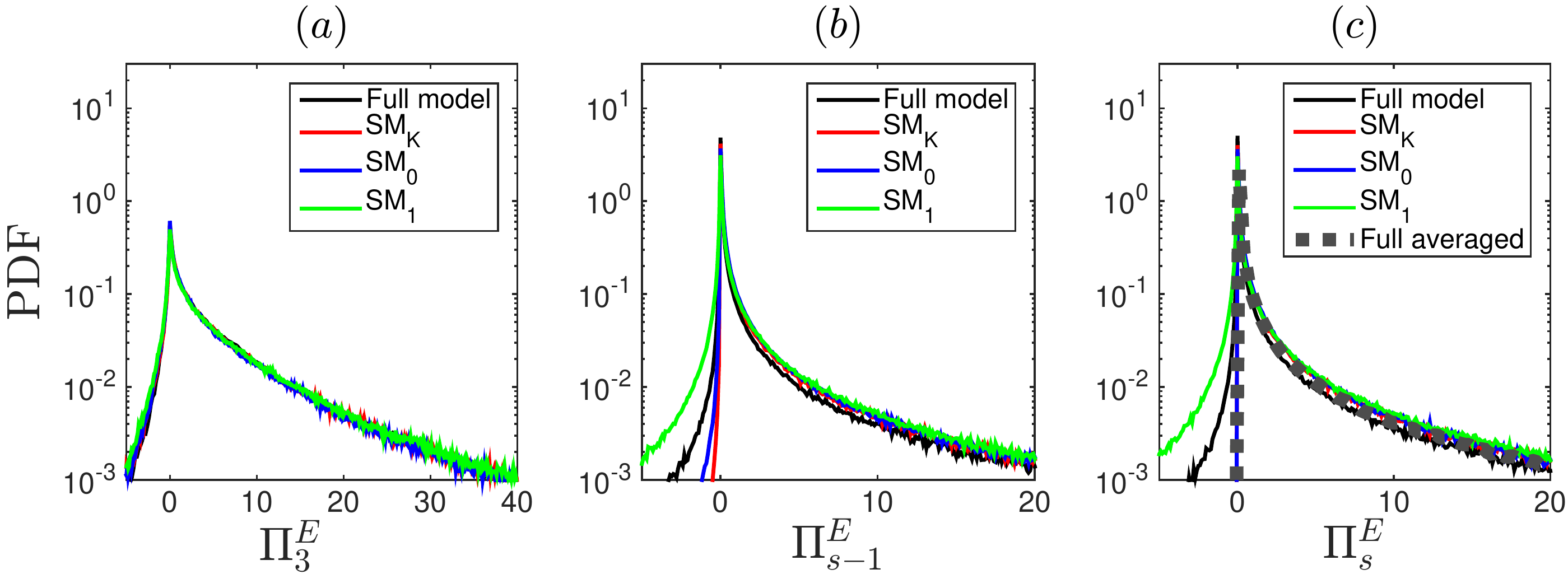}
\caption{PDFs for the energy flux at shell (a) $n = 3$, (b) $n = s-1$ and (c) $n = s$. Black line is the result of the full viscous model. Red, blue and green lines correspond, respectively, to the subgrid models SM${}_K$, SM${}_0$ and SM${}_1$ with $s = 20$ shells. Bold grey dotted line represents the distribution for the full model, averaged with respect to the shells $n > s$.}
\label{figB6}
\end{figure}

The time-averaged values of the energy and helicity fluxes computed at shell corresponding to the small-scale range $n \ll s$ are in very good agreement 
with the unclosed model. 
For example, in  Fig.~\ref{figB6} we compare the PDFs of energy fluxes for different models, computed at the large-scale shell $n = 3$ and for the last two shells $s-1 = 19$ and $s = 20$. One can see that a very good convergence is attained for large scales, but again the some  discrepancies are observed for the small scales. Here the models SM${}_K$ and SM${}_0$ show no energy backscattering events with a  strictly positive flux as predicted by Eq.~(\ref{eqExtra2b}). On the other hand, the full model and the model SM${}_1$ show energy backscattering events (negative energy flux). 

For interpreting this result, it is necessary to recall that the subgrid models were designed for the statistical distribution averaged over the shell numbers $n > s$. Thus, the reference quantity for checking the validity of subgrid models must be the energy flux averaged in the same manner. We performed such an averaging of the energy flux numerically for the full viscous model: {using the long-time simulation results, the flux function $\Pi_s^E$ was averaged over the points with nearby values of shell speeds $u_s$ and $u_{s-1}$ but arbitrary $u_{s+1}$ and $u_{s+2}$. The PDF for the resulting (conditionally averaged) flux is shown in Fig.~\ref{figB6}(c) by bold grey dotted line. This results shows a rather surprising result that, after averaging over shells $n > s$, the energy flux $\Pi_s^E$ becomes strictly positive.} This also means that the models SM$_K$ and SM$_0$ have the correct behavior, while the backscattering events demonstrated by the model SM${}_1$ (green line in Fig.~\ref{figB6}(c)) should not be interpreted in favor of this subgrid model.

It is also instructive to compute the flatness in terms of the energy flux as $$
F_n^\Pi = \frac{\langle |\Pi_n^E|^{4/3} \rangle}{\langle |\Pi_n^E|^{2/3} \rangle^2}.$$
 The corresponding numerical results are given in Fig.~\ref{figN6} (c), demonstrating a slightly more regular behavior. Deviations are still present at cutoff shells, but this time the model SM${}_0$ shows a better match with the full model.

{\subsection{Why improved subgrid models do not work better?} }

In subgrid models discussed in Sections~\ref{sec:num1} and \ref{sec:numtest1}, we used direct averages for equations of the last shells, or the average conditioned on one multiplier $z_s$. Though the closures constructed in this way are rather simplistic, the very fast decay of correlations of multipliers with a shell separation suggests that even such simple models should be reasonably accurate, see Fig.~\ref{fig1} (c,d). On the contrary, we saw that more accurate models do not demonstrate any improvement for the statistics at small scales. In order to verify this observation with higher-order approximations, we also constructed the subgrid model that depends on two multipliers, $z_s$ and $z_{s-1}$. This was done by using expansions in multi-dimensional spherical harmonics, which is possible due to a homogeneity property of $\mathcal{A}_1$ and $\mathcal{A}_2$ as functions of shell speeds $(u_s,u_{s-1},u_{s-2})$. Details of this study are rather lengthy and will not be presented here. The conclusion is, however, the same: no considerable improvement was observed for the statistics at small scales,  in comparison with the simplest model SM${}_K$. In this section we propose a possible explanation to this unfortunate lack of convergence.

The distribution limited to a finite number of shells in the inertial region, $P(u_1,\ldots,u_s)$, can be expected to be a smooth positive function, as one can infer from Kolmogorov's third hypothesis (see Section~\ref{sec4}). Our subgrid models are constructed with the purpose of recovering this regular distribution in a statistical sense, i.e., as an attractor. However, the regularity of the distribution imposes a very strong requirement on the subgrid model, which is reminiscent to the mixing property for measure-preserving dynamical systems: an infinitely long trajectory of the subgrid model must be dense everywhere in the configuration space. As we know from the dynamical system theory, such a property is structurally unstable for a non-conservative dynamical system, like our subgrid model. This implies that an arbitrarily small change of the ``ideal'' subgrid system may drastically change its long-time statistical behavior; see also \cite{berkooz1994observation}, where this issue was investigated for the Lorenz system. In a general case, one can expect that the subgrid model possesses a chaotic (fractal) attractor, therefore, occupying only a zero measure subset in configuration space.  This structural instability may be the main cause of the persistent divergence from the full model statistics in our subgrid models. It is important to notice that this high sensitivity to the closure is nevertheless limited to a fixed (Reynolds independent) number of shells, indicating that the large scale dynamics is {\it robust} and {\it universal} with respect to the small-scale closure, i.e. we do not have a strong sensitivity of the global attractor  on the fine details of high frequency fluctuations.

\begin{figure}
\centering
\includegraphics[width = 0.8\textwidth]{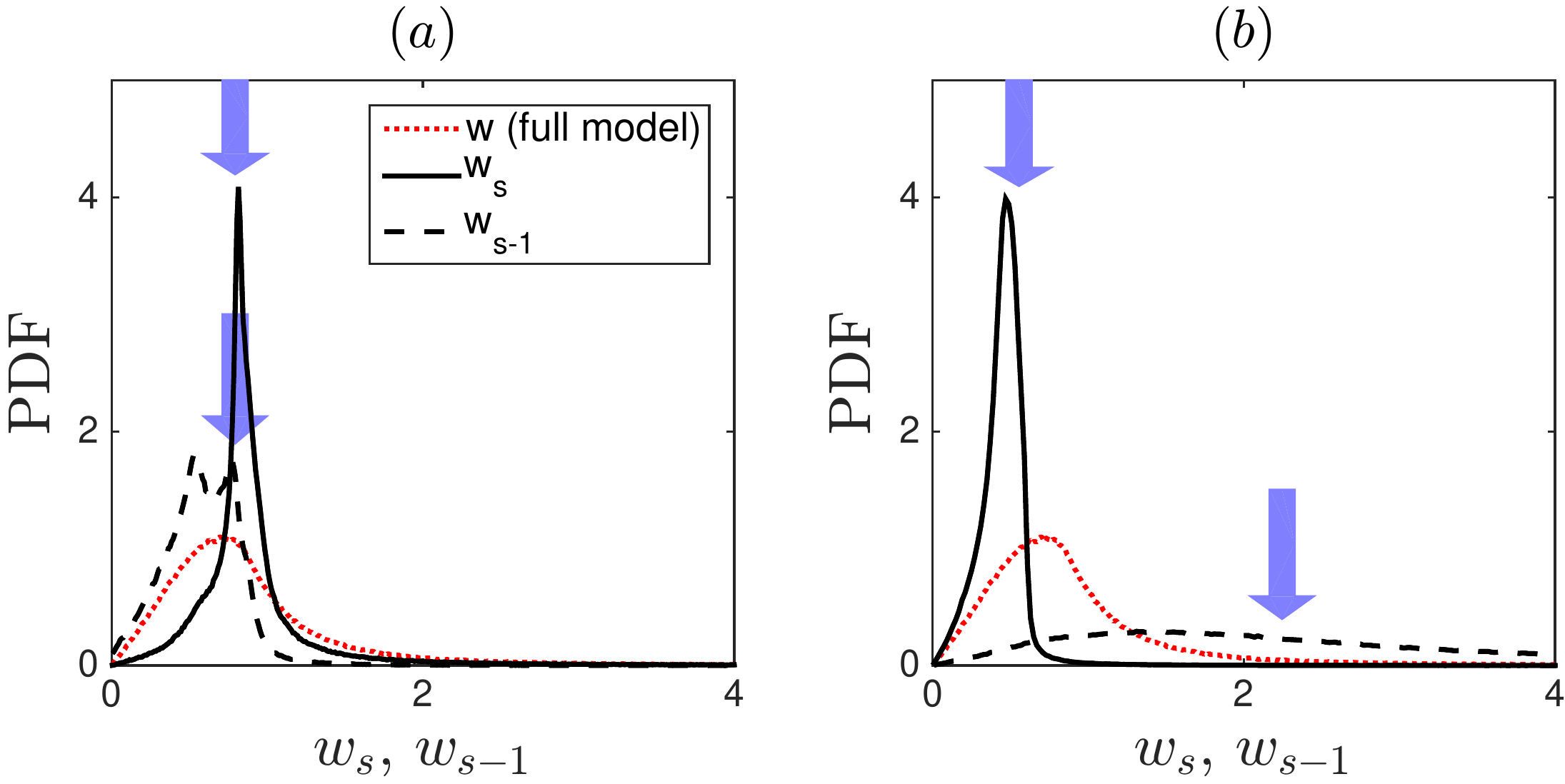}
\caption{PDFs for the absolute values of multipliers, $w_s$ (solid line) and $w_{s-1}$ (dashed line), for the models (a) SM${}_K$ and (b) SM${}_0$. {Arrows mark positions of the constant multipliers of the time-independent solution. Dotted red line indicates the universal PDF of the full model.}
}
\label{figB7}
\end{figure}

It is possible to demonstrate some quantitative evidence in favor of our hypothesis. The inviscid Sabra model has unstable time-independent solutions of Kolmogorov type, which up to phase-symmetry factors have the form 
\begin{equation}
u_n = ia_nk_n^{-1/3}, \quad a_n = a_{n+3},
\label{extraF1}
\end{equation}
with arbitrary period-3 real coefficients $a_n$, see e.g.~\cite{biferale1995transition,mailybaev2016spontaneous}. 
In this case, all multipliers (\ref{eqSM3}) are fixed numbers, $z_n = -i\lambda^{-1/3}a_n/a_{n-1}$, similarly to the Desnyansky--Novikov shell model in Section~\ref{sec4b}.
For our subgrid models, the factors $a_n$ are not arbitrary and can be found by substituting (\ref{extraF1}) into the equations $R_{s-1} = 0$ and $R_s = 0$, see~(\ref{eqE7x}) and (\ref{eqE17}). The elementary computations with expressions (\ref{eqTN9}) yield
\begin{equation}
\frac{a_s}{a_{s-1}} = \lambda^{-2/3}\frac{\mathcal{A}_1}{\mathcal{A}_2},\quad
\frac{a_{s-1}}{a_{s-2}} = \lambda^{1/3}\frac{\mathcal{A}_2}{i\mathcal{A}_1^2}.
\label{extraF2}
\end{equation}
For the two simplest models, where $\mathcal{A}_{1,2}$ are constants (see Section~\ref{sec:num1}), we have
\begin{align}
\textrm{model SM}_K:\quad &\frac{a_s}{a_{s-1}} = 1,\quad
\frac{a_{s-1}}{a_{s-2}} = -1,
\label{extraF3}
\\[5pt]
\textrm{model SM}_0:\quad &\frac{a_s}{a_{s-1}} \approx 0.7,\quad 
\frac{a_{s-1}}{a_{s-2}} \approx -2.83.
\label{extraF4}
\end{align}
Figure~\ref{figB7} shows the PDFs for the absolute values of multipliers, $w_s = |z_s|$ and $w_{s-1} = |z_{s-1}|$, in these two models obtained from numerical simulations. Blue arrows indicate the positions given by the time-independent solutions (\ref{extraF1})--(\ref{extraF4}) as $w_s = \lambda^{-1/3}|a_s/a_{s-1}|$ and $w_{s-1}= \lambda^{-1/3}|a_{s-1}/a_{s-2}|$. One can clearly see the correlation between the PDFs and time-independent solutions, which can be explained by the intermittent dynamics alternating between the chaotic and regular behavior, see, e.g., \cite{ott2002chaos}. One can also see some footprints of this temporary ``locking'' to a time-independent solution for
 the model SM${}_K$ in Fig.~\ref{figN2} (f), where a small plateau is developed with slowly changing amplitudes. 
 
\begin{figure}
\centering
\includegraphics[width = 0.8\textwidth]{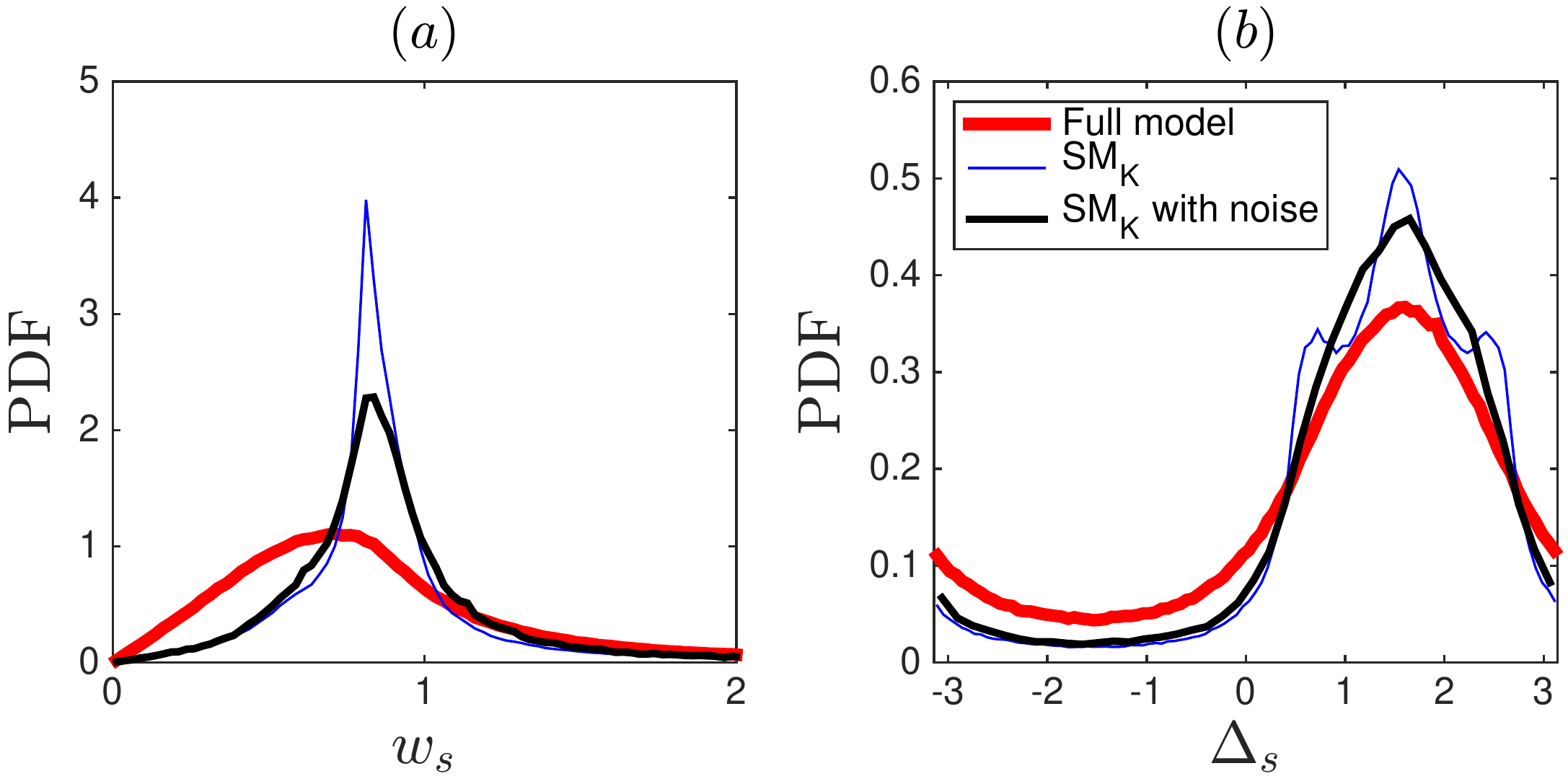}
\caption{{PDFs for (a) multipliers $w_s$ and (b) phases $\Delta_{s}$. The results of the full model (bold red) are compared with the model SM${}_K$ (thin blue line) and the same model after addition of noise (black line).}
}
\label{figB8}
\end{figure}

We conclude that the chaotic attractor may be influenced by the time-independent solution at small scales. As we explained, this defect is generic and, hence, hardly can be removed by using more accurate subgrid models, {unless some special extra conditions are imposed. For example,
noise can be used as a mechanisms to improve the dynamics at subgrid scales, by counteracting the attraction to specific solutions. In order to see how large is the effect, we applied random phase perturbations $\Delta_{s+1} = \pi/2+x_1$ and $\Delta_{s+2} = \pi/2+x_2$ in the models SM$_K$, see Section~\ref{sec:num1}. Here $x_1(t)$ and $x_2(t)$ are obtained as solutions of the Langevin equation $\dot{x} = -\tau^{-1}x+\sigma\tau^{-1/2}\xi $ with the Kolmogorov time scale $\tau = k_s^{-2/3}$ and white noise $\xi$ with $\langle\xi(t)\rangle = 0$ and $\langle\xi(t)\xi(t')\rangle = \delta(t-t')$. To have a moderate noise level we choose $\sigma = 0.4$. Figure~\ref{figB8} shows the PDFs of the multipliers $w_s$ and phases $\Delta_s$ at the last shell of the model with $s = 20$. One can see that the noise has some effect improving the distributions, as compared with the bold red PDFs of the full model, but this effect is small. The results suggest that adding an uncorrelated noise to the evolution of phase  is not an effective mechanism for our subgrid model.}

\section{Conclusions}
\label{sec:conc}

We have discussed a theoretical framework to define the optimal sub-grid closure for shell models of turbulence. 
The theoretical framework would predict a very complicated sub-grid models which depends on the conditional probability of sub-grid variables on all resolved scales, a task unrealistic even for the simple structure of shell models.
We have proposed a series of approximate  closures  based on the ansatz that consecutive  shell {\it multipliers} are short-range correlated, following the third hypothesis of Kolmogorov formulated for similar quantities for the  three-dimensional Navier--Stokes  turbulence. Different  approximations assume different degrees of correlations across scales among amplitudes and phases of consecutive multipliers. We show numerically that such low order closures work well, reproducing all known properties of the large-scale dynamics including anomalous scaling. We found small but systematic discrepancies only for a range of scales close to the sub-grid model, which do not tend to disappear by increasing the order of the approximation. We speculate that the lack of convergence might be due to a breaking of ergodicity at least for the evolution of very fast degrees of freedom at small scales. Effects of the sub-grid closure on the resolved range of scales must be quantified  also for real LES of three-dimensional Navier--Stokes equations.  Correlations between the sub-grid stress tensor and velocity increments at the resolved scales can be estimated on the basis of fusion-rules \cite{lvo96a,ben98}. They  are supposed to be sub-leading with respect to the scaling of single-scale velocity increments \cite{meneveau94}, i.e. reproducing the same kind of sensitivity to the particular closure only for a range of separations close 
 to the sub-grid cutoff. A quantitative assessment of the importance of such a feedback is nevertheless missing. 

{\it Acknowledgments.} 
The authors are grateful to R. Benzi, M. Cencini and G.L. Eyink for useful comments. 
The work was supported by the European Research Council under the European Union's Seventh Framework Programme, ERC Grant Agreement No 339032. 
A.A.M. was also supported by the CNPq (Grant No. 302351/2015-9) and FAPERJ (grant E-26/210.874/2014). 

\bibliography{biblio}

\end{document}